\documentclass[pre, aps, showpacs, showkeys]{revtex4}
\usepackage{graphicx}
\begin{document}

\title{Lagrangean formulation of ion- and dust-ion-acoustic waves
\footnote{Proceedings of the \textit{International Conference on
Plasma Physics - ICPP 2004}, Nice (France), 25 - 29 Oct. 2004;
contribution P3-045; available online at:
\texttt{http://hal.ccsd.cnrs.fr/ccsd-00001891/en/}  .}}

\author{I. Kourakis\footnote{On leave from: U.L.B. - Universit\'e Libre
de Bruxelles, Physique Statistique et Plasmas C. P. 231, Boulevard
du Triomphe, B-1050 Brussels, Belgium; also: Facult\'e des
Sciences Apliqu\'ees - C.P. 165/81 Physique G\'en\'erale, Avenue
F. D. Roosevelt 49, B-1050 Brussels, Belgium;
\\Electronic address: \texttt{ioannis@tp4.rub.de}} and P. K.
Shukla\footnote{Electronic address: \texttt{ps@tp4.rub.de}}}
\affiliation{Institut f\"ur Theoretische Physik IV,
Fakult\"at f\"ur Physik und Astronomie, \\
Ruhr--Universit\"at Bochum, D-44780 Bochum, Germany}

\date{\today}

\begin{abstract}
Ion-acoustic modes propagating in unmagnetized dusty plasmas are
studied by applying a generic collisionless fluid model. An
Eulerian-to-Lagrangean variable transformation leads to a new
system of evolution equations, which may be combined into a single
(rather complex) equation for the mean ion density. The linear and
weakly nonlinear oscillations regimes are studied and their
modulational stability is investigated. This study extends known
previous results on the Lagrangian description of electron- or
ion-acoustic electrostatic plasma modes. The relevance with the
description of dust-ion acoustic waves propagating in a dusty
plasma is also discussed.
\end{abstract}

\pacs{52.25.Lw, 52.35.Fp, 52.35.Mw, 52.35.Sb}

\keywords{Electrostatic waves, ion--acoustic mode, nonlinear
waves.}

\maketitle

\section{Introduction}

Dusty plasmas (DP) have recently attracted a great deal of
attention due to a variety of new phenomena observed in them and
the novel physical mechanisms involved in their description
\cite{PSbook, Verheest}. In addition to known plasma electrostatic
waves \cite{Krall}, e.g. ion-acoustic waves (IAW), new oscillatory
modes arise in DP \cite{PSbook, Verheest}, among which the
\textit{dust-ion acoustic wave} (DIAW), essentially corresponding
to ion-acoustic (IA) oscillations in a strong presence of dust.

The linear properties of the IAWs and DIAWs have been quite
extensively studied and now appear well understood \cite{PSbook}.
As far as nonlinear effects are concerned, various studies have
pointed out the possibility of the formation of DIAW-related
localized structures, due to a mutual compensation between
nonlinearity and dispersion, including small-amplitude pulse
solitons, shocks and vortices \cite{PSsolitons}. Furthermore, the
propagation of nonlinearly modulated DIA wave packets was studied
in Ref. \cite{IKPSDIAW}, in addition to the formation of localized
envelope soliton--modulated waves due to the modulational
instability of the carrier waves.  A very interesting known
approach is the Lagrangean description of a nonlinear  wave
profile. In the context of electrostatic plasma waves, this
formalism has been employed in studies of electron plasma waves
\cite{Davidson1, Davidson2, Infeld} and, more recently,
ion-acoustic \cite{Chakra1} and dust-acoustic \cite{Chakra2}
waves. Our aim here is to extend previous results by applying the
Lagrangean formalism to the description of nonlinear DIAWs
propagating in dusty plasmas.

\section{The model}

We shall consider the nonlinear propagation of dust-ion-acoustic
waves in a collisionless plasma consisting of three distinct
particle species `$\alpha$': an inertial species of ions (denoted
by `$i$'; mass $m_i$, charge $q_i = + Z_i e$; $e$ denotes the
absolute of the electron charge), surrounded by an environment of
thermalized electrons (mass $m_e$, charge $- e$), and massive dust
grains (mass $M$, charge $q_d = s Z_d e$, both assumed constant
for simplicity); $Z_d$ denotes the charge state of dust grains; we
leave the choice of dust grain charge sign $s = q_d/|q_d|$ ($=
-1/+1$ for negative/positive dust charge) open in the algebra.
Charge neutrality is assumed at equilibrium.

The hydrodynamic--Poisson system of equations which describe the
evolution of the ion `fluid' in the plasma are considered. The ion
number density $n_i$  is governed by the continuity equation
\begin{equation}
\frac{\partial n_i}{\partial t} + \nabla  (n_i \,{\bf u}_i)= 0 \,
, \label{densityequation}
\end{equation}
where the mean velocity ${\bf u}_i$ obeys
\begin{equation}
\frac{\partial {\bf u}_i}{\partial t} + {\bf u}_i  \cdot \nabla
{\bf u}_i \, = \, \frac{Z_i e}{m_i} {\bf E} \, = \, - \frac{Z_i
e}{m_i}\,\nabla \,\Phi \, .
\end{equation}
The electric field ${\bf E} = -\nabla \,\Phi $ is related to the
gradient of the wave potential $\Phi$, which is obtained from
Poisson's equation $\nabla \cdot {\bf E} = 4 \pi \, \sum q_s\, n_s
$, viz.
\begin{equation}
\nabla^2 \Phi \,
= \,4 \pi \,e \,(n_c  \,+ n_h  \, - Z_i \,n_i)  \, .
\label{Poisson}
\end{equation}
Alternatively, one may consider
\begin{equation}
\frac{\partial E}{\partial t} \, =\, - 4 \pi \, \sum_\alpha
q_\alpha\, n_\alpha \,u_\alpha  \, . \label{Poisson2}
\end{equation}
 We assume a near-Boltzmann distribution for the electrons,
 i.e. \(n_e \approx n_{e,0}\, \exp(e \Phi/k_B T_e) \,
\) ($T_e$ is the electron temperature and $k_B$ is Boltzmann's
constant). The dust distribution is assumed stationary, i.e. $n_d
\approx {\rm{const.}}$. The overall quasi-neutrality condition at
equilibrium then reads
\begin{equation}
Z_i \, n_{i, 0} \, + s Z_d \, n_{d} \,  -  n_{e, 0} \,  =\, 0 \, .
\label{neutrality}
\end{equation}

\subsection{Reduced Eulerian equations \label{reducedEuler}}

By choosing appropriate physical scales, Eqs. (1)-(3) can be cast
into a reduced (dimensionless) form. Let us define the
ion-acoustic speed $c_{s} = (k_B T_{e}/m_i)^{1/2}$. An appropriate
choice for the space and timescales, $L$ and $T = L/c_{s}$, are
the effective Debye length $\lambda_{D} = (k_B T_{e}/4 \pi Z_i^2
n_{i, 0} e^2)^{1/2} \equiv c_{s}/\omega_{p, i}$ and the  ion
plasma period $\omega_{p, i}^{-1} = (4 \pi n_{i, 0} Z_i^2
e^2/m_i)^{- 1/2}$, respectively.  Alternatively, one might leave
the choice of $L$ (and thus $T = L/c_{s}$) arbitrary -- following
 an idea suggested in Refs. \cite{Chakra1, Chakra2}) -- which
 leads to the appearance of a dimensionless dispersion parameter
$\delta = 1/(\omega_{p, i} T) = \lambda_{D}/L$ in the formulae.
The specific choice of scale made above corresponds to $\delta =
1$ (implied everywhere in the following, unless otherwise stated);
however, we may keep the parameter $\delta$
 to `label' the dispersion term in the forthcoming formulae.

For one-dimensional wave propagation along the $x$ axis, Eqs.
(\ref{densityequation}) - (\ref{Poisson}) can now be written as
\begin{eqnarray}
\frac{\partial n}{\partial t} + \frac{\partial (n \, u)}{\partial
x} & = & 0\, ,
\nonumber \\
\frac{\partial u}{\partial t} + u  \frac{\partial u}{\partial x}
\, & = & \,- \nabla \phi \,
,\nonumber \\
\delta^2 \, \frac{\partial^2 \phi}{\partial x^2} \, & = & \,
\,(\hat n - n)\, , \label{reducedeqs}
\end{eqnarray}
where all quantities are dimensionless: $n = n_i/n_{i, 0}$,
$\mathbf{u} = \mathbf{u}_i/v_{0}$ and $\phi = \Phi/\Phi_0$; the
scaling quantities are, respectively: the equilibrium ion density
$n_{i, 0}$, the effective sound speed $v_{0} = c_{s}$ (defined
above) and $\Phi_{0} = k_B T_{e}/(Z_i e)$. The (reduced) electron
and dust background density $\hat n$ is defined as
\begin{equation}
\hat n = \frac{n_{e}}{Z_i n_{i, 0}}\, e^{\phi/Z_i} + s \frac{Z_d
n_{d}}{Z_i n_{i, 0}}\, \equiv \mu \, e^{\phi/Z_i} + 1 - \mu \, ,
\label{defhatn1}
\end{equation}
where we have defined the DP parameter $\mu = {n_{e, 0}}/({Z_i
n_{i, 0}})$, and made use of Eq. (\ref{neutrality}). Note that
both $n$ and $\hat n$ reduce to unity at equilibrium.

We shall define, for later reference,  the function $f(\phi) =
\hat n$ -- given by Eq. (\ref{defhatn1}) -- and its inverse
function
\begin{equation}
f^{-1}(x) = Z_i \, \ln \biggl( 1 + \frac{x - 1}{\mu} \biggr)
\equiv g(x) \, , \label{defg} \end{equation} viz. $f(\phi) = x$
implies $\phi = f^{-1}(x) \equiv g(x)$.

We note that the dependence on the charge sign $s$ is now
incorporated in $\mu = 1 + s {Z_d n_{d, 0}}/({Z_i n_{i, 0}})$;
retain that $\mu < 1$ ($\mu > 1$) corresponds to negative
(positive) dust. Remarkably, since the dust-free limit is
recovered for $\mu = 1$, the results to be obtained in the
following are also straightforward valid for ion-acoustic waves
propagating in (dust-free) e-i plasma, upon setting $\mu = 1$ in
the formulae.

The well--known DIAW dispersion relation $\omega^2 = c_{s}^2
k^2/(k^2 \lambda_{D}^2 + 1)$ \cite{IKPSDIAW} is obtained from Eqs.
(\ref{densityequation}) to (\ref{neutrality}). On the other hand,
the system (\ref{reducedeqs}) yields the reduced relation
$\omega^2 = k^2/(\delta^2 k^2  + 1)$, which of course immediately
recovers the former dispersion relation upon restoring dimensions
(regardless, in fact, of one's choice of space scale $L$; cf.
definition of $\delta$). However, some extra qualitative
information is admittedly hidden in the latter (dimensionless)
relation. Should one consider a very long space scale $L \gg
\lambda_D$ (i.e. $\delta \ll 1$), one readily obtains $\omega \sim
k$ (unveiling the role of $\delta$ as a characteristic dispersion
control parameter). Finally, the opposite limit of short $L$ (or
infinite $\delta$) corresponds to ion plasma oscillations (viz.
$\omega = \omega_{p, i}$ = constant).

\subsection{Lagrangean description}

Let us introduce the Lagrangean variables $\{ \xi, \tau \}$, which
are related to the Eulerian ones $\{ x, t \}$ via
\begin{equation}
\xi \, = \, x\, - \int_0^\tau u(\xi ,\tau')\, d\tau' \, , \qquad
\qquad \tau = t \, . \label{Lagrange-def}
\end{equation}
See that they coincide at $t = 0$. Accordingly, the space and time
gradients are transformed as
\[
\partial/\partial x \rightarrow \alpha^{-1} \, \partial/\partial \xi \,
, \qquad
\partial/\partial t \rightarrow \partial/\partial \tau -
\alpha^{-1} \, u \, \partial/\partial \xi\, , \] where we have
defined the quantity
\begin{equation}
\alpha(\xi, \tau) \equiv \frac{\partial x}{\partial \xi} = 1 +
\int_0^\tau d\tau' \frac{\partial }{\partial \xi} u(\xi ,\tau') \,
. \label{L0}
\end{equation}
Note that the convective derivative $D \equiv \partial/\partial t
+ u \,\partial/\partial x$ is now plainly identified to
$\partial/\partial \tau$.  Also notice that $\alpha$ satisfies
$\alpha(\xi, \tau = 0) = 0$ and
\begin{equation}
\frac{\partial \alpha(\xi, \tau)}{\partial \tau} = \frac{\partial
u(\xi, \tau)}{\partial \xi} \label{property1}
\end{equation}
As a matter of fact, the Lagrangean transformation defined here
reduces to a Galilean transformation if one suppresses the
evolution of $u$, i.e. for $u = {\rm const.}$ (or $\partial
u/\partial \tau =
\partial u/\partial \xi = 0$, hence $\alpha = 1$).
Furthermore, if one also suppresses the dependence in time $\tau$,
this transformation is reminiscent of the travelling wave ansatz
$f(x, t) = f(x - v t \equiv s)$, which is widely used in the
Sagdeev potential formalism \cite{PSsolitons, Sagdeev}.

The Lagrangean variable transformation defined above leads to a
new set of reduced equations
\begin{eqnarray}
n(\xi, \tau) &=& \alpha^{-1}(\xi, \tau) \, n(\xi, 0) \label{L1} \\
\frac{\partial u(\xi, \tau)}{\partial \tau} &=& \frac{Z_i e}{m_i}
E(\xi, \tau) \nonumber \\
&=& - \frac{Z_i e}{m_i} \,\alpha^{-1}(\xi, \tau) \,\frac{\partial
\phi(\xi, \tau)}{\partial \xi}  \qquad \label{L2} \\
\alpha^{-1}(\xi, \tau) \,\frac{\partial E(\xi, \tau)}{\partial
\xi} &=& 4 \pi
Z_i e [n(\xi, \tau) - \hat n \,n_{i, 0}] \label{L3} \\
\biggl( \frac{\partial }{\partial \tau} - \alpha^{-1} u \,
\frac{\partial }{\partial \xi} \biggr) E(\xi, \tau) &=& - 4 \pi
Z_i e n(\xi, \tau) u(\xi, \tau) \, , \label{L4}
\end{eqnarray}
where we have temporarily restored dimensions for physical
transparency; recall that the (dimensionless) quantity $\hat n $,
which is in fact a function of $\phi$, is given by
(\ref{defhatn1}). One immediately recognizes the role of the
(inverse of the) function $\alpha(\xi, \tau)$ as a density time
evolution operator; cf. Eq. (\ref{L1}) \cite{comment1}. Poisson's
equation is now obtained by eliminating $\phi$ from Eqs.
(\ref{L2}, \ref{L3})
\begin{equation}
\alpha^{-1} \,\frac{\partial }{\partial \xi} \biggl( \alpha^{-1}
\, \frac{\partial \phi}{\partial \xi} \biggr) \,= - 4 \pi Z_i e (n
- \hat n \,n_{i, 0}) \, . \label{L5}
\end{equation}
Note that a factor $\delta^2$ should appear in the left-hand side
if one rescaled Eq. (\ref{L5}) as described above; cf. the last of
Eqs. (\ref{reducedeqs}). This will be retained for later
reference, with respect to the treatment suggested in Ref.
\cite{Chakra1} (see discussion below).

 In principle, our aim is to solve the system of Eqs.
(\ref{L1}) to (\ref{L4}) or, by eliminating $\phi$, Eqs.
(\ref{L1}), (\ref{L2}) and (\ref{L5}) for a given initial
condition $n(\xi, \tau=0) = n_0(\xi)$, and then make use of the
definition (\ref{Lagrange-def}) in order to invert back to the
Eulerian arguments of the state moment variables (i.e. density,
velocity etc.). However, this abstract scheme is definitely not a
trivial  task to accomplish.

\section{Nonlinear dust-ion acoustic oscillations}

Multiplying Eq. (\ref{L3}) by $u(\xi, \tau)$ and then adding to
Eq. (\ref{L4}), one obtains
\begin{equation}
\frac{\partial E(\xi, \tau)}{\partial \tau} = - 4 \pi Z_i e n_{i,
0}\, \hat n \, u(\xi, \tau) \label{L6} \, .
\end{equation}
Combining with Eq. (\ref{L2}), one obtains
\begin{equation}
\frac{\partial^2 u}{\partial \tau^2} = - \omega_{p, i}^2 \, \hat n
\, u \, ,\label{NLoscil}
\end{equation}
where $\omega_{p, i}$ is the ion plasma frequency (defined above).
Despite its apparent simplicity, Eq. (\ref{NLoscil}) is
{\em{neither}} an ordinary differential equation (ODE) -- since
all variables depend on {\em{both}} time $\tau$ and space $\xi$ --
{\em{nor}} a closed evolution equation for the mean velocity
$u(\xi, \tau)$: note that the (normalized) background particle
density $\hat n$ depends on the potential $\phi$ and on the plasma
parameters; see its definition (\ref{defhatn1}). The evolution of
the potential $\phi(\xi, \tau)$, in turn, involves $u(\xi, \tau)$
(via the quantity $\alpha(\xi, \tau)$) and the ion density $n(\xi,
\tau)$.

Eq. (\ref{NLoscil}) suggests that the system performs nonlinear
oscillations at a frequency $\omega = \omega_{p, i} \, {\hat
n}^{1/2}$. Near equilibrium, the quantity ${\hat n}$ is
approximately equal to unity and one plainly recovers a linear
oscillation at the ion plasma frequency $\omega_{p, i}$. Quite
unfortunately this apparent simplicity, which might in principle
enable one to solve for $u(\xi, \tau)$ and then obtain $\{ \xi,
\tau \}$ in terms of $\{ x, t \}$ and vice versa (cf. Davidson's
treatment for electron plasma oscillations in Ref.
\cite{Davidson2}; also compare to Ref. \cite{Infeld}, setting
$\gamma = 0$ therein), is absent in the general (off-equilibrium)
case where the plasma oscillations described by Eq.
(\ref{NLoscil}) are intrinsically {\em{nonlinear}}.

Since Eq. (\ref{NLoscil}) is in general not a closed equation for
$u$, unless the background density $\hat n$ is constant (i.e.
independent of $\phi$, as in Refs.  \cite{Davidson2, Infeld}), one
can \emph{neither} apply standard methods involved in the
description of nonlinear oscillators on Eq. (\ref{NLoscil}) (cf.
Ref. \cite{Infeld}), \emph{nor} reduce the description to a study
of Eqs. (\ref{NLoscil}, \ref{L6}) (cf. Ref. \cite{Davidson1}), but
rather has to retain all (or rather five) of the evolution
equations derived above, since five inter-dependent dynamical
state variables (i.e. $n$, $u$, $E$, $\phi$ and $\alpha$) are
involved. This procedure will be exposed in the following Section.

\section{Perturbative nonlinear Lagrangean treatment}

Let us consider weakly nonlinear oscillations performed by our
system close to (but not at) equilibrium. The basis of our study
will be the reduced system of equations
\begin{eqnarray}
\frac{\partial }{\partial \tau} (\alpha \, n) = 0 \, , \nonumber \\
\frac{\partial u}{\partial \tau} = E \, , \nonumber \\
\frac{\partial E}{\partial \xi} = (n - \hat n)\, \alpha \, ,
\nonumber
\\
\alpha \, E = - \frac{\partial  \phi}{\partial \xi} \, , \nonumber \\
\frac{\partial \alpha}{\partial \tau} = \frac{\partial u}{\partial
\xi} \, , \label{system-reduced}
\end{eqnarray}
which follow from the Lagrangean Eqs. (\ref{L1}) to (\ref{L5}) by
scaling over appropriate quantities, as described in \S
\ref{reducedEuler} \cite{comment2}. This system describes the
evolution of the state
 vector, say ${\mathbf{S}} = (\alpha, n, u, E, \phi)$ ($\in \Re^5$), in
the
 Lagrangean coordinates defined above. We will consider small
deviations
 from the equilibrium state ${\mathbf{S}_0} = (1, 1, 0, 0, 0)^T$, by
 taking ${\mathbf{S}} = {\mathbf{S}}^{(0)} + \epsilon
{\mathbf{S}_1}^{(0)} +
 \epsilon^2 {\mathbf{S}_2}^{(0)} + ...$, where $\epsilon$ ($\ll 1$) is
a
 smallness parameter.  Accordingly, we shall Taylor develop the
quantity
 $\hat n(\phi)$ near $\phi \approx 0$, viz. $\phi \approx \epsilon
\phi_1 +
 \epsilon^2 \phi_2 + ...$, in order to express $\hat n$ as
 \begin{eqnarray}
 \hat n & \approx & 1 + c_1 \phi + c_2 \phi^2 + c_3 \phi^3 + ...
 \nonumber \\
 & = &  1 + \epsilon c_1 \phi_1 + \epsilon^2
 (c_1 \phi_2 + c_2 \phi_1^2)  + \epsilon^3
 (c_1 \phi_3 + 2 c_2 \phi_1 \phi_2 + c_3 \phi_1^3) + ... \, ,
 \end{eqnarray}
 where the coefficients $c_j$ ($j=1, 2, ...$), which are determined
from
 the definition (\ref{defhatn1}) of $\hat n$,
 contain all the essential dependence on the plasma parameters,
 e.g. $\mu$;
 making use of $e^{x} \approx \sum_{n=0}^\infty {x}^n/n!$,
 one readily obtains
 \[
 c_1 = \mu/Z_i \, , \qquad c_2 = \mu/(2 Z_i^2)  \, , \qquad c_2 =
\mu/(6 Z_i^3)
 \, . \]
Remember that for $\mu = 1$ (i.e.  for vanishing dust) one
recovers the expressions for IAWs in e-i plasma.

Following the standard reductive perturbation technique
\cite{redpert}, we shall consider the stretched (slow) Lagrangean
coordinates \( Z \,= \, \epsilon (\xi - \lambda \,\tau) \, , \quad
T \,= \, \epsilon^2 \, \tau\) (where $\lambda \in \Re$ will be
determined later). The perturbed state of (the $j-$th --- $j = 1,
..., 5$ --- component of) the state vector  ${\mathbf{S}}^{(n)}$
is assumed to depend on the fast scales via the carrier phase
$\theta = k \xi - \omega \tau$, while the slow scales enter the
argument of the ($j-$th element's) $l-$th harmonic amplitude
$S_{j, l}^{(n)}$, viz. \( S{j}^{(n)} \,= \,
\sum_{l=-\infty}^\infty \,S_{j, l}^{(n)}(Z, \, T)
 \, e^{i l (k \xi - \omega \tau)}\) (where $S_{j,-l}^{(n)} =
 {S_{j, l}^{(n)}}^*$ ensures reality).
Treating the derivative operators as
\[
\frac{\partial}{\partial \tau} \rightarrow
\frac{\partial}{\partial \tau} - \epsilon \, \lambda \,
\frac{\partial}{\partial Z} + \epsilon^2 \,
\frac{\partial}{\partial T} \, , \qquad \frac{\partial}{\partial
\xi} \rightarrow \frac{\partial}{\partial \xi} + \epsilon \,
\frac{\partial}{\partial Z} \, ,
\]
and substituting into the system of evolution equations, one
obtains an infinite series in both (perturbation order)
$\epsilon^n$ and (phase harmonic) $l$. The standard perturbation
procedure now consists in solving in successive orders $\sim
\epsilon^n$ and substituting in subsequent orders. The method
involves a tedious calculation which is however  straightforward;
the details of the method are presented e.g. in Ref.
\cite{IKPSDIAW}, so only the essential stepstones need to be
provided here.

The equations obtained for $n = l = 1$ determine the first
harmonics of the perturbation
\begin{equation}
n_1^{(1)} \, = - \alpha_1^{(1)} \, = \, ({k^2}/{\omega^2})  \psi
\, , u_1^{(1)} \, = (k/\omega) \psi \, , \qquad E_1^{(1)} \, = - i
k \psi \label{1st-order-corrections}
\end{equation}
where $\psi$ denotes the potential correction $\phi_1^{(1)}$. The
cyclic frequency $\omega$ obeys the dispersion relation \(
\omega^2\,  = {k^2}/({k^2 + s c_1})\), which exactly recovers,
once dimensions are restored, the standard IAW dispersion relation
\cite{Krall} mentioned above.

Proceeding in the same manner, we obtain the second order
quantities, namely the amplitudes of the second harmonics
$\mathbf{S}_2^{(2)}$ and constant (`direct current') terms
$\mathbf{S}_0^{(2)}$, as well as a finite contribution
$\mathbf{S}_1^{(2)}$ to the first harmonics; as expected from
similar studies, these three (sets of 5, at each $n, l$)
quantities are found to be proportional to $\psi^2$, $|\psi|^2$
and $\partial \psi/\partial Z$ respectively; the lengthy
expressions are omitted here for brevity. The ($n = 2$, $l=1$)
equations provide the compatibility condition: \( \lambda \,=
\omega (1 - \omega^2)/k = {d \omega}/{d k}\); $\lambda$ is
therefore the group velocity $v_g(k) = \omega'(k)$ at which the
wave envelope propagates. It turns out that $v_g$ decreases with
increasing wave number $k$; nevertheless, it always remains
positive.

In order $\sim \epsilon^3$, the equations for $l = 1$ yield an
explicit compatibility condition in the form of a nonlinear
Schr\"odinger--type equation (NLSE)
\begin{equation}
i\, \frac{\partial \psi}{\partial T} + P\, \frac{\partial^2
\psi}{\partial Z^2} + Q \, |\psi|^2\,\psi = 0 \, .  \label{NLSE}
\end{equation}
Recall that $\psi\, \equiv \,  \phi_1^{(1)}$ denotes the amplitude
of the first-order electric potential perturbation. The `slow'
variables $\{ Z, T \}$ were defined above.

The {\em dispersion coefficient} $P$ is related to the curvature
of the dispersion curve as \( P \,  = \, \omega''(k)/{2} \,= - 3
\omega^3 (1- \omega^2)/(2 k^2)\). One may easily check that $P$ is
negative (for all values of $k$).

 The {\em nonlinearity coefficient} $Q$ is due to carrier
wave self-interaction. It is given by the expression
\begin{equation}
Q = + \frac{\omega^3}{12 \, k^4} \frac{\mu}{Z_i^4}\, \biggl[ 3
Z_i^3 k^6 - 3 (\mu + 4) Z_i^2 k^4
+ 3 (1 - 2 \mu - 5 \mu^2) Z_i k^2 - \mu (3 \mu - 1)^2 \biggr] \, .
 \label{Qcoeff}
\end{equation}
where the coefficients $c_{1, 2, 3}$ were defined above.

For low wavenumber $k$, $Q$ goes to $- \infty$ as
 \[
Q \approx - \frac{(3 \mu - 1)^2 \mu^{1/2}}{12 \, Z_i^{5/2}} \,
\frac{1}{k} \, .
\]

\subsection{Modulational stability analysis}

According to the standard analysis \cite{Hasegawa}, we can
linearize around the plane wave solution of the NLSE (\ref{NLSE})
\(\psi \, = \, {\hat \psi} \, e^{i Q |\hat \psi|^2 \tau} \, + \,
c.c. \, , \) ($c.c.$: complex conjugate) -- notice the amplitude
dependence of the frequency shift $\Delta \omega = \epsilon^2 Q
|\hat \psi|^2$ -- by setting \({\hat \psi} \, = \, {\hat \psi}_0
\, + \, \epsilon \, {\hat \psi}_1 \, , \) and then assuming the
perturbation ${\hat \psi}_1$ to be of the form: ${\hat \psi}_1 \,
= \, {\hat \psi}_{1, 0} \,e^{i ({\hat k} \zeta - {\hat \omega}
\tau)} \, + \, c.c.$. Substituting into (\ref{NLSE}), one thus
readily obtains \( \hat \omega^2 \, = \, P^2 \, \hat k^2 \,
\biggl(\hat k^2 \, - \, 2 ({Q}/{P}) |\hat \psi_{1, 0}|^2 \biggr)
\). The wave will thus be {\em stable} ($\forall \, \hat k$) if
the product $P Q$ is negative. However, for positive $P  Q > 0$,
instability sets in for wavenumbers below a critical value $\hat
k_{cr} = \sqrt{2 Q/P}\, |\hat \psi_{1, 0}|$, i.e. for wavelengths
above a threshold $\lambda_{cr} = 2 \pi/\hat k_{cr}$; defining the
instability growth rate \( \sigma = |Im\hat\omega(\hat k)| \), we
see that it reaches its maximum value for $\hat k = \hat
k_{cr}/\sqrt{2}$, viz.
\[ \sigma_{max} =
|Im\hat\omega|_{\hat k = \hat k_{cr}/\sqrt{2}} \,=\, | Q |\, |\hat
\psi_{1, 0}|^2  \, .
\]
We see that the instability condition depends only on the sign of
the product $P Q$, which may be studied numerically, relying on
the expressions derived above.

\subsection{Finite amplitude nonlinear excitations}

The NLSE (\ref{NLSE}) is long known to possess distinct types of
localized constant profile (solitary wave) solutions, depending on
the sign of the product $P Q$ \cite{Hasegawa, Fedele, IKPSDIAW}.
Remember that this equation here describes the evolution of the
wave's envelope, so these solutions represent slowly varying
localized envelope structures, confining the (fast) carrier wave.
The analytic form of these excitation can be found in the
literature (see e.g. in \cite{IKPSDIAW} for a brief review) and
need not be derived here in detail. Let us however briefly
summarize those results.

Following Ref. \cite{Fedele}, we may seek a solution of Eq.
(\ref{NLSE}) in the form \( \psi(\zeta, \tau) = \rho(Z, T) \,
e^{i\,\Theta(\zeta, \tau) } + {\rm c.c.}\),  where $\rho$,
$\sigma$ are real variables which are determined by substituting
into the NLSE and separating real and imaginary parts. The
different types of solution thus obtained are summarized in the
following.

\begin{figure}[!]
 \centering
 \resizebox{5in}{!}{
\includegraphics{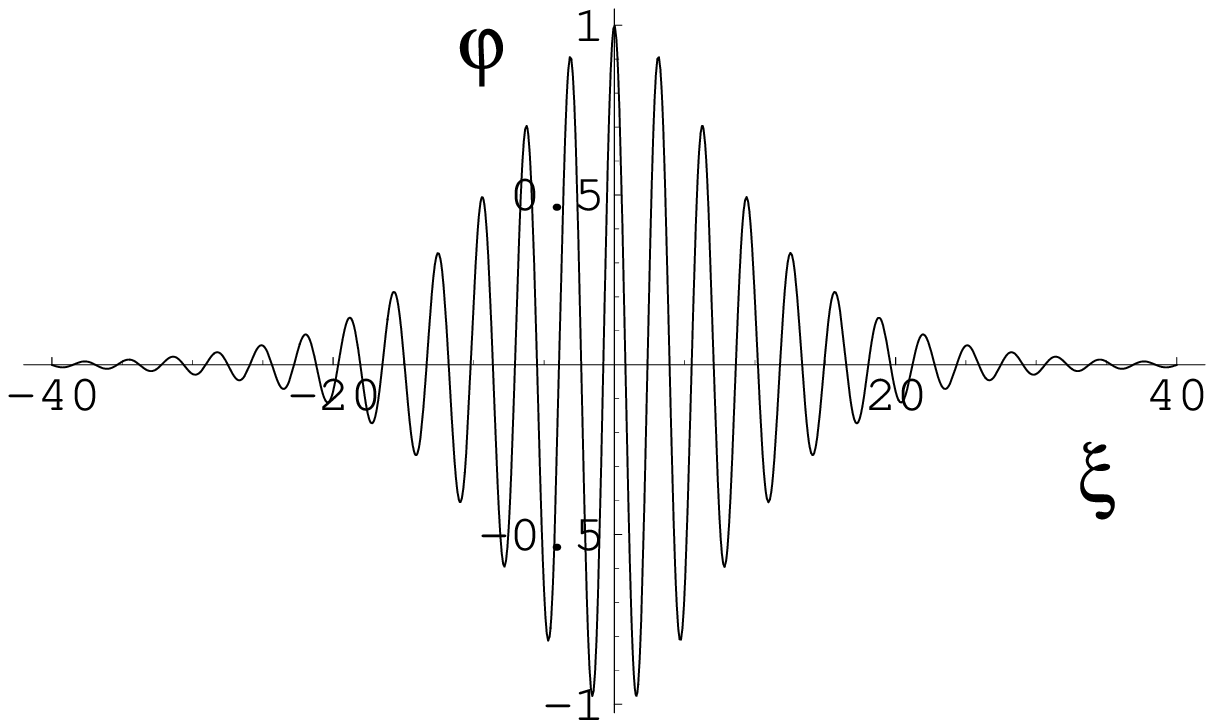}
\hskip 1 cm
\includegraphics{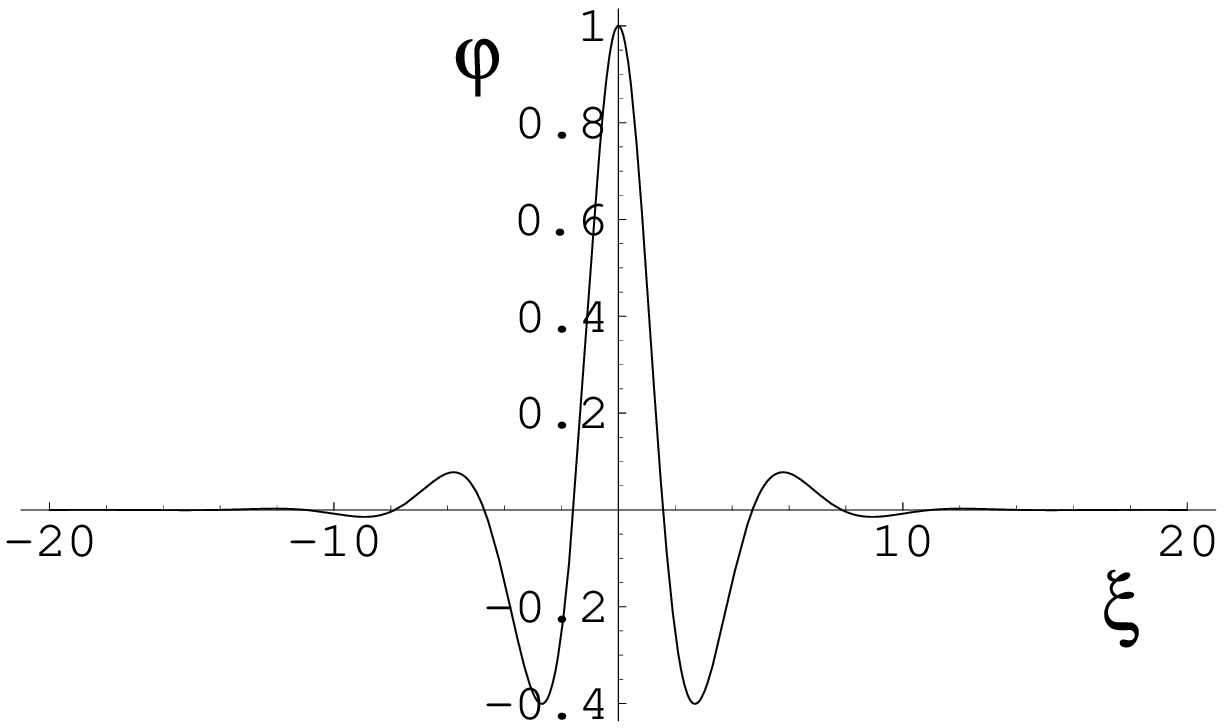}
}
\\
 \resizebox{5in}{!}{
 \includegraphics[]{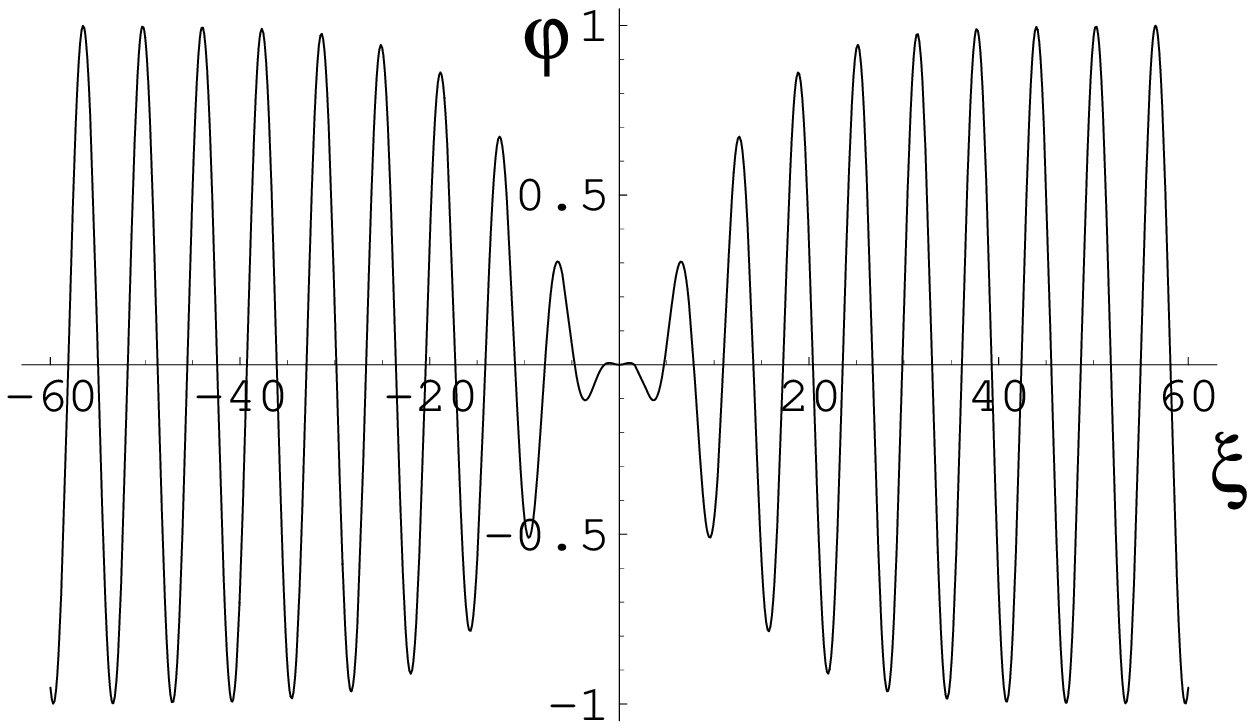}\hskip 1 cm
\includegraphics{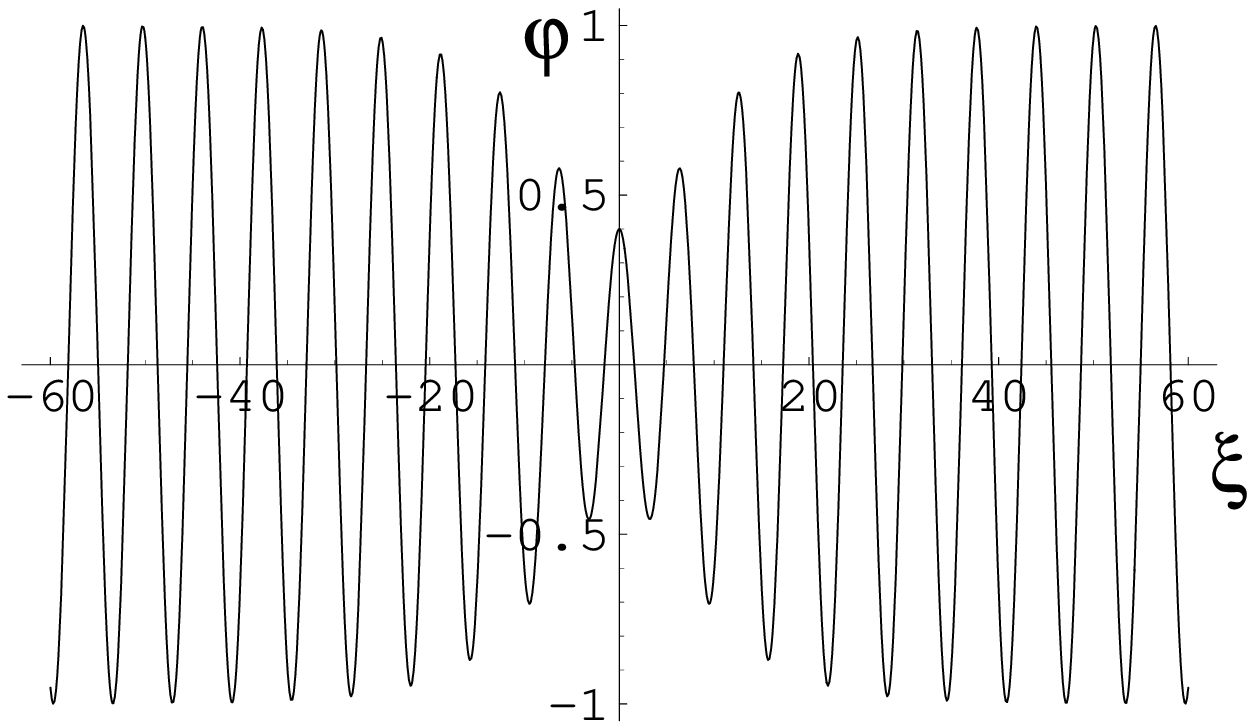}
} \caption{A heuristic representation of wave packets modulated by
solutions of the NLS equation. These envelope excitations are of
the: (a, b) bright type ($P Q > 0$, pulses); (c) dark type, (d)
gray type ($P Q < 0$, voids). Notice that the amplitude never
reaches zero in (d).} \label{figure1}
\end{figure}

For $P Q > 0$ we find the {\em (bright) envelope soliton}
\begin{equation}
\rho = \pm \rho_0 \, sech \biggl(\frac{Z - u_e\, \tau}{L} \biggr)
\, , \quad  \Theta = \frac{1}{2 P} \, \bigl[ u_e Z - (\Omega +
\frac{1}{2} u_e^2) T \bigr] \, , \label{bright}
\end{equation}
which represents a localized pulse travelling at the envelope
speed $u_e$ and oscillating at a frequency $\Omega$ (at rest). The
pulse width $L$ depends on the maximum amplitude square $\rho_0$
as \( L = ({2 P}/{Q })^{1/2}/\rho_0 \). Since the product $P Q$ is
always positive for long wavelengths, as we saw above, this type
of excitation will be rather privileged in dusty plasmas. The
bright-type envelope soliton is depicted in Fig. \ref{figure1}a,
b.

For $P Q < 0$, we obtain the {\em dark} envelope soliton ({\em
hole}) \cite{Fedele}
\begin{eqnarray}
\rho & = & \pm \rho_1 \, \biggl[ 1 -  sech^2 \biggl(\frac{Z - u_e
T}{L'} \biggr)\biggr]^{1/2}   \pm \rho_1 \,
 \tanh \biggl(\frac{Z - u_e\, T}{L'}
\biggr) \, ,
\nonumber \\
\Theta & = & \frac{1}{2 P} \, \biggl[ u_e Z \, -
\biggl(\frac{1}{2} u_e^2 - 2 P Q \rho_1 \biggr) \,T \biggr] \, ,
\label{darksoliton}
\end{eqnarray}
which represents a localized region of negative wave density
(shock) travelling at a speed $u_e$; see Fig. \ref{figure1}c.
Again, the pulse width depends on the maximum amplitude square
$\rho_1$ via \( L' = (2 \bigl|{P}/{Q}\bigr|)^{1/2}/\rho_1 \).

Finally, still for $P Q < 0$, one also obtains the {\em gray}
envelope solitary wave \cite{Fedele}
\begin{equation}
\rho = \pm \rho_2 \, \biggl[ 1 - a^2\, sech^2 \biggl(\frac{Z - u_e
T}{L''} \biggr)\biggr]^{1/2} \, , \label{greysoliton}
\end{equation}
which also represents a localized region of negative wave density.
Comparing to the dark soliton (\ref{darksoliton}), we note that
the maximum amplitude $\rho_2$ is now finite (non-zero)
everywhere; see Fig. \ref{figure1}d. The
 the pulse width of this gray-type excitation
\( L'' = \sqrt{2 | {P}/{Q} |}/(a \,\rho_2) \) now also depends on
an independent parameter $a$ which represents the modulation depth
($0 < a \le 1$). The lengthy expressions which determine the phase
shift $\Theta$ and the parameter $a$, which are omitted here for
brevity, can be found in Refs. \cite{Fedele, IKPSDIAW}. For $a =
1$, one recovers the {\em dark} soliton presented above.

An important qualitative result to be retained is that the
envelope soliton width $L$ and maximum amplitude $\rho$ satisfy $L
\rho \sim \sqrt{P/Q}$ (see above), and thus depend on (the ratio
of) the coefficients $P$ and $Q$; for instance, regions with
higher values of $P$ (or lower values of $Q$) will support wider
(spatially more extended) localized excitations, for a given value
of the maximum amplitude. Contrary to the KdV soliton picture, the
width of these excitations does not depend on their velocity. It
does, however, depend on the plasma parameters, e.g. here $\mu$.

\begin{figure}[htb]
 \centering
 \resizebox{3in}{!}{
\includegraphics{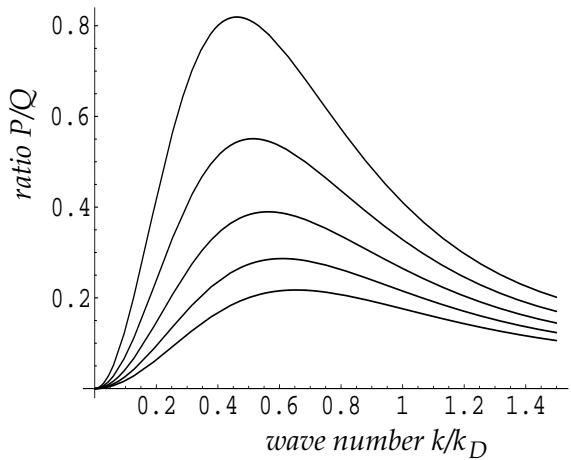}
} \caption{The ratio $\eta = P/Q$ of the coefficients in the NLSE
(\ref{NLSE}) is depicted versus the wave number $k$ (normalized
over $k_D$), for several values of the dust parameter $\mu$; in
descending order (from top to bottom): 0.8, 0.9, 1.0, 1.1, 1.2. }
\label{figure2}
\end{figure}

The localized envelope excitations presented above represent the
slowly varying envelope which confines the (fast) carrier space
and time oscillations, viz. $\phi = \Psi(X, Z) \cos(k \xi - \omega
\tau)$ for the electric potential $\phi$ (and analogous
expressions for the density $n_i$ etc.; cf.
(\ref{1st-order-corrections})). The qualitative characteristics
(width, amplitude) of these excitations, may be investigated by a
numerical study of the ratio $P/Q \equiv \eta(k; \mu)$: recall
that its sign determines the type (bright or dark) of the
excitation, while its (absolute) value determines its width for a
given amplitude (and vice versa). In Fig. \ref{figure2} we have
depicted the behaviour of $\eta$ as a function of the wavenumber
$k$ and the parameter $\mu$: higher values of $\mu$ correspond to
lower curves. Remember that, for any  given wavenumber $k$, the
dust concentration (expressed via the value of $\mu$) determines
the soliton width $L$ (for a given amplitude $\rho$; see
discussion above) since $L \sim \eta^{1/2}/\rho$. Therefore, we
see that the addition of {\em{negative}}  dust generally ($\mu <
1$) results to higher values of $\eta$ (i.e. wider or higher
solitons), while {\em{positive}} dust ($\mu > 1$) has the opposite
effect: it reduces the value of  $\eta$ (leading to narrower or
shorter solitons). In a rather general manner, bright type
solitons (pulses) seem to be rather privileged, since the ratio
$\eta$ (or the product $P Q$) of the coefficients $P$ and $Q$ is
positive in most of the $k, \mu$ plane of values. One exception
seems to be very the region of {\em{very}} low values of $\mu$
(typically below $0.2$), which develops a negative tail of $\eta$
for small $k$ ($< 0.3 \, k_D$): thus, a very high ($> 80$ per
cent) electron depletion results in pulse destabilization in
favour of dark-type excitations (Fig. \ref{figure1}c, d). Strictly
speaking, $\eta$ also becomes negative for very high wave number
values ($> 2.5 \, k_D$); nevertheless, we neglect -- for rigor --
this region from the analysis, in this (long wavelength $\lambda$)
fluid picture (for a weak dust presence, short $\lambda$ DIAWs may
be quite strongly damped; however, this result may still be
interesting for a strong presence of dust, when Landau damping is
not a significant issue \cite{PSbook}).

\section{Relation to previous works:
an approximate nonlinear Lagrangean treatment}

By combining the Lagrangean system of Eqs. (\ref{L1}) to
(\ref{L5}), one obtains the (reduced) evolution equation
\begin{equation}
\frac{\partial^2}{\partial \tau^2} \biggl( \frac{1}{n}  \biggr) =
- \frac{1}{n_0} \frac{\partial }{\partial \xi} \, \biggl[
\frac{n}{n_0} \,\frac{\partial }{\partial \xi} \, g(w) \biggr] \,
, \label{Lfinal}
\end{equation}
where the function $g(x)$ [defined in Eq. (\ref{defg})] is
evaluated at
\[w(n) = n \biggl[1 - \delta^2 \frac{\partial^2}{\partial \tau^2}
\biggl( \frac{1}{n} \biggr) \biggr] \, .\] Note that the ion
density $n$ has been scaled by its equilibrium value $n_{i, 0}$,
to be distinguished from the initial condition $n_0 = n(\xi, \tau
= 0)$.

Despite its complex form, the nonlinear evolution equation
(\ref{Lfinal}) can be solved exactly by considering different
special cases, as regards the order of magnitude of the
dispersion--related parameter $\delta$. This treatment, based on
Ref. \cite{Chakra1}, will only be briefly summarized here, for the
sake of reference.


First, one may consider very short scale variations, i.e. $L \ll
\lambda_D$ (or $\delta \gg 1$). This amounts to neglecting
collective effects, so oscillatory motion within a Debye sphere is
essentially decoupled from neighboring ones. By considering \(w(n)
\approx - \delta^2 \, n \, {\partial^2}\bigl( {1}/{n}
\bigr)/{\partial \tau^2} \) and $\phi \approx 0$ (i.e. $\hat n
\approx 1$), one may combine Eqs. (\ref{L4}) and (\ref{Lfinal})
into
\begin{equation}
\biggl( \frac{\partial^2 }{\partial \tau^2} + \omega_{p, i}^2
\biggr) \biggl( \frac{1}{n} - 1 \biggr) = 0 \, ,
\end{equation}
which, imposing the initial condition $n(\xi, 0) = n_0(\xi)$,
yields the solution
\begin{equation}
n(\xi, \tau) = \frac{n_0(\xi)}{\frac{n_0(\xi)}{n_{i, 0}} + \bigl(1
- \frac{n_0(\xi)}{n_{i, 0}} \bigr) \cos \omega_{p, i}\tau} \, .
\end{equation}
Note that if the system is initially at equilibrium, viz.
$n_0(\xi) = n_{i, 0}$, then it remains so at all times $\tau > 0$.
Now, one may go back to Eq. (\ref{L1}) and solve for $\alpha(\xi,
\tau)$, which in turn immediately provides the mean fluid velocity
$u$
\[
u(\xi, \tau) = \omega_{p, i} \sin \omega_{p, i}  \tau
\int_{\xi_0}^\xi \biggl(1 - \frac{n_0(\xi')}{n_{i, 0}} \biggr)
d\xi'
\]
via (\ref{L0}), and then $E(\xi, \tau)$ and $\phi(\xi, \tau)$.
Finally, the variable transformation (\ref{Lagrange-def}) may now
be inverted, immediately providing the Eulerian position $x$ in
terms of $\xi$ and $\tau$. We shall not go into further details
regarding this procedure, which is essentially analogue (yet not
identical) to Davidson's treatment of electron plasma
oscillations.


Quite interestingly, upon neglecting the dispersive effects, i.e.
setting $\delta = 0$, Eq. (\ref{Lfinal}) may be solved by
separation of variables, and thus shown to possess a nonlinear
special solution in the form of a product, say $n(\xi,\tau) =
n_1(\xi) n_2(\tau)$ \cite{comment2}. This calculation was put
forward in Ref. \cite{Chakra1} (where the study of IAW -- in a
single electron temperature plasma -- was argued to rely on an
equation quasi-identical to Eq. (\ref{Lfinal})). However, the
solution thus obtained relies on doubtful physical grounds, since
the assumption $\delta \approx 0$, which amounts to remaining
close to equilibrium -- cf. the last of Eqs. (\ref{reducedeqs}),
implies an infinite space scale $L$ (recall the definition of
$\delta$), contrary to the very nature of the (localized)
nonlinear excitation itself. Rather not surprisingly, this
solution was shown in Ref. \cite{Chakra1} to decay fast in time,
in both Eulerian and Lagrangean coordinates. Therefore, we shall
not pursue this analysis any further.

\section{Discussion and conclusions}

We have studied the nonlinear propagation of dust ion acoustic
waves propagating in a dusty plasma.  By employing a Lagrangean
formalism, we have investigated the modulational stability of the
amplitude of the propagating dust ion acoustic oscillations and
have shown that these electrostatic waves may become unstable, due
to self interaction of the carrier wave.  This instability may
either lead to wave collapse or to wave energy localization, in
the form of propagating localized envelope structures.  We have
provided an exact set of analytical expressions for these
localized excitations.

This study complements similar investigations which relied on an
Eulerian formulation of the dusty plasma fluid model
\cite{IKPSDIAW}. In fact, the Lagrangean picture provides a
strongly modified nonlinear stability profile for the wave
amplitude, with respect to the previous (Eulerian) description;
this was intuitively expected, since the passing to Lagrangean
variables involves an inherently nonlinear transformation, which
inevitably modifies the nonlinear evolution profile of the system
described. However, the general qualitative result remains in
tact: the dust ion acoustic-type electrostatic plasma waves may
propagate in the form of localized envelope excitations, which are
formed as a result of the mutual balance between dispersion and
nonlinearity in the plasma fluid. More sophisticated descriptions,
incorporating e.g. thermal or collisional effects, may be
elaborated in order to refine the parameter range of the problem,
and may be reported later.

\medskip

\begin{acknowledgments}
This work was funded by the {\it{SFB591 (Sonderforschungsbereich)
-- Universelles Verhalten gleichgewichtsferner Plasmen: Heizung,
Transport und Strukturbildung}} German government Programme.
Support by the European Commission (Brussels) through the Human
Potential Research and Training Network via the project entitled:
``Complex Plasmas: The Science of Laboratory Colloidal Plasmas and
Mesospheric Charged Aerosols'' (Contract No. HPRN-CT-2000-00140)
is also acknowledged.
\end{acknowledgments}


\begin{thebibliography}{10}

\bibitem{PSbook} P. K. Shukla and A. A. Mamun, \textit{Introduction to
Dusty Plasma Physics} (Institute of Physics Publishing Ltd.,
Bristol, 2002).

\bibitem{Verheest} F. Verheest, \textit{Waves in Dusty Space Plasmas}
(Kluwer Academic Publishers, Dordrecht, 2001).

\bibitem{Krall} N. A. Krall and A. W. Trivelpiece,
\textit{Principles of plasma physics}, McGraw - Hill (New York,
1973); Th. Stix, \textit{Waves in Plasmas}, American Institute of
Physics (New York, 1992).

\bibitem{PSsolitons} For a review, see:
P. K. Shukla and A. A. Mamun, New J. Phys. \textbf{5}, 17.1
(2003).

\bibitem{IKPSDIAW} I.Kourakis and P. K. Shukla,
\textit{Physics of Plasmas} \textbf{10} (9), 3459 (2003);
\textit{Eur. Phys. J. D} \textbf{28}, 109 (2003).

\bibitem{Davidson1} R. C. Davidson and P. P. J. M. Schram,
\textit{Nuclear Fusion} \textbf{8}, 183 (1968).

\bibitem{Davidson2} R. C. Davidson, {\it{Methods in nonlinear plasma
theory}}, Academic Press (New York, 1972).

\bibitem{Infeld}
E. Infeld and G. Rowlands, Phys. Rev. Lett. {\textbf{58}} (1987).

\bibitem{Chakra1} N. Chakrabarti and M. S. Janaki,
{\it Phys. Lett.} A {\bf 305} 393 (2002).

\bibitem{Chakra2} N. Chakrabarti and M. S. Janaki,
{\it Phys. Plasmas} {\bf 10} 3043 (2003).

\bibitem{Sagdeev} R. Z. Sagdeev, in {\it Reviews of Plasma Physics},
Vol. 4., ed. M. A. Leontovich, Consultants Bureau (New York,
1966), p.52.

\bibitem{comment1} Eq. (\ref{L1}) was obtained from the
(Lagrangean) density equation, which is reduced to: $\partial(n
\alpha)/\partial \tau = 0$ by using the property
(\ref{property1}); Eq. (\ref{L1}) then follows.

\bibitem{comment2} Eqs. (\ref{system-reduced}) are derived
from Eqs. (\ref{L1}, \ref{L2}, \ref{L3}), $E=-\nabla \phi$ and
(\ref{property1}), respectively. We have avoided the appearance of
$\alpha^{-1}$ -- cf. Eqs. (\ref{L4}, \ref{L5}) -- for analytical
convenience.

\bibitem{redpert} T. Taniuti and N. Yajima, \, J. Math. Phys. {\bf 10},
1369 (1969); N. Asano,\, T. Taniuti and \, N. Yajima, \, J. Math.
Phys. {\bf 10}, 2020 (1969).

\bibitem{Hasegawa} A. Hasegawa,
\textit{Plasma Instabilities and Nonlinear Effects}
(Springer-Verlag, Berlin, 1975).

\bibitem{Fedele} R. Fedele, H. Schamel and P. K. Shukla,
{\it Phys. Scripta} T {\bf 98} 18 (2002); R. Fedele and H.
Schamel, {\it Eur. Phys. J. B} {\bf 27} 313 (2002); Fedele, {\it
Phys. Scripta} {\bf 65} 502 (2002).

\end{thebibliography}
\end{document}